\documentclass[aps,prd,superscriptaddress,showpacs,nofootinbib,notitlepage,twocolumn]{revtex4-1}
\usepackage{graphicx,subfigure,amsmath,amsfonts,amssymb,multirow,mathrsfs}
\usepackage[colorlinks,linkcolor=blue,citecolor=blue,urlcolor=blue]{hyperref}
\usepackage[dvipsnames,svgnames]{xcolor}
\usepackage{amsmath}

\begin{document}

\title{A new group of low-spin $50-70M_\odot$ Black Holes and the high pair-instability mass cutoff}

\author{Yuan-Zhu Wang$^{\dagger}$}
\affiliation{Institute for Theoretical Physics and Cosmology, Zhejiang University of Technology, Hangzhou, 310032, People's Republic of China}

\author{Yin-Jie Li$^{\dagger}$}
\affiliation{Key Laboratory of Dark Matter and Space Astronomy, Purple Mountain Observatory, Chinese Academy of Sciences, Nanjing 210023, People's Republic of China}

\author{Shi-Jie Gao}
\affiliation{School of Astronomy and Space Science, Nanjing University, Nanjing, 210023, People's Republic of China}

\author{Shao-Peng Tang}
\affiliation{Key Laboratory of Dark Matter and Space Astronomy, Purple Mountain Observatory, Chinese Academy of Sciences, Nanjing 210023, People's Republic of China}

\author{Yi-Zhong Fan}
\email{The corresponding author: yzfan@pmo.ac.cn\\$^\dagger$Contributed equally}
\affiliation{Key Laboratory of Dark Matter and Space Astronomy, Purple Mountain Observatory, Chinese Academy of Sciences, Nanjing 210023, People's Republic of China}
\affiliation{School of Astronomy and Space Science, University of Science and Technology of China, Hefei, Anhui 230026, People's Republic of China}

\begin{abstract}
Pair-instability supernovae (PISN) will not leave compact remnants and hence yield a mass gap of the black holes. 
Though a transition point at 
$\approx 46M_\odot$, separating low- and high-spin black hole populations and interpreted as evidence for the PISN mass gap, was first identified in gravitational wave data by Wang et al. (2022, ApJL 941, L39) and later confirmed in follow-up studies, here we report the emergence of a {\it new} group of low-spin but massive ($\sim 50-70M_\odot$) black holes, which are hard to produce via hierarchical mergers, in the latest GWTC-4.0 data. Correspondingly, the mass cutoff of the low-spin black holes shifts to $68.5^{+19.8}_{-18.5}M_\odot$ (90\%  credibility), which is consistent with the PISN model for a $^{12}{\rm C}(\alpha,\gamma)^{16}{\rm O}$ reaction rate of $S_{300} = 109^{+55}_{-27}~{\rm keV~b}$.  
Despite that the massive single-star collapse/dynamical capture origin can not be reliably tested at this moment, 
a high pair-instability mass cutoff $M_{\rm low}\sim 70M_\odot$ may be favored for its capability of accounting for the rather low observation rate of hydrogen-less super-luminous supernovae.
\end{abstract}


\maketitle

\section{Introduction} \label{sec:intro}
The direct detection of gravitational waves (GWs) from binary black hole (BBH) mergers by Advanced LIGO and Virgo has inaugurated a new era of observational astrophysics \cite{2016PhRvL.116f1102A}. This breakthrough has provided an unprecedented tool for testing general relativity in the strong-field regime \cite{2025PhRvD.112h4080A,2025PhRvL.135k1403A,2026SciBu..71...83T} and, crucially, for studying the population of stellar-mass BHs, which are otherwise hardly visible to electromagnetic observatories \cite{2021NatAs...5..749G,2022PhR...955....1M}. One of the key goals of the population studies of the BBH events is to find evidence for the (pulsational) pair-instability supernovae (PISN \cite{1967PhRvL..18..379B}) that are predicted to leave no BHs within the mass range of $(M_{\rm low},~M_{\rm high})$, where $M_{\rm low}\sim 40-90 {M_\odot}$ and $M_{\rm high}\sim 120-175 {M_\odot}$ \cite{2016A&A...594A..97B, 2017MNRAS.470.4739S, 2017ApJ...836..244W,2020ApJ...902L..36F,2021ApJ...912L..31W}, commonly referred to as the pair-instability mass gap (PIMG).
The existence, edges, and detailed structure of this gap are affected by factors such as the $^{12}{\rm C}(\alpha,\gamma)^{16}{\rm O}$ reaction rate, as well as the properties of massive stellar progenitors (e.g., metallicity, rotation, and mass-loss rates), among which the reaction rate has the greatest influence \cite{2020ApJ...902L..36F,2021ApJ...912L..31W,2022ApJ...924...39M}. 

The first gravitational-wave transient catalog (GWTC-1) resulted in a preliminary hint for a cutoff in the primary mass distribution at $\sim 45 {M_\odot}$ \cite{2019PhRvX...9c1040A}, which however has been challenged by the more massive black holes identified in subsequent observations. 
Considering the plausible presence of a high-mass BH component resulting from hierarchical mergers \cite{2020PhRvL.125j1102A,2021NatAs...5..749G,2021ApJ...915L..35K}, Wang et al. found that a cutoff feature at $m_{\rm c}\approx 50M_\odot$ could exist in the GWTC-2 data \cite{2021ApJ...913...42W}, and interpreted it as $M_{\rm low}$, in view of its consistency with stellar evolution models (see also \cite{2021ApJ...916L..16B}). 
Motivated by such progress, in 2022 Wang et al. \cite{2022ApJ...941L..39W} further constructed
phenomenological models containing two subpopulations, and found out that the black holes in GWTC-3 can indeed be well described by two components characterized by their different mass/spin features. 
In the dedicated analysis focusing on spin magnitudes, it turns out that the spin amplitude of component BHs can be divided into two groups with a division mass $m_{\rm d}=46.1^{+5.6}_{-5.1}M_\odot$ \cite{2022ApJ...941L..39W}, 
comparable to the cutoff mass $m_{\rm c}$ in \cite{2021ApJ...913...42W}, 
and has been taken as the signature of the PISN. This two-subpopulation scenario in which the low-mass/spin sub population truncates a at $\approx 45M_\odot$ has been 
further supported by Li et al. \cite{2024PhRvL.133e1401L} using a flexible mixture model for the component-mass versus spin-magnitude distribution of GWTC-3 BHs.
Owing to their robust analysis, those authors interpreted this as evidence for the pair-instability explosions of massive stars. Subsequent studies with the same dataset have obtained similar results using different approaches
\cite{2024ApJ...977...67L,2024arXiv240601679P,2024ApJ...975...54G,2025PhRvL.134a1401A,2025ApJ...987...65L,2025arXiv250819208M}.

Very recently, the GWTC-4.0 data are available and the BBH event sample has increased significantly \cite{2025arXiv250818082T}. 
With these data, Tong et al. \cite{2025arXiv250904151T} found a cut-off mass of $45^{+5}_{-4} {M_\odot}$ in the spectrum of secondary mass (see however Ref.\cite{2025arXiv251018867R} for different opinion with a more flexible mass function). Antonini et al. \cite{2025arXiv250904637A} reported a rapid transition at $45.3^{+6.5}_{-4.8} {M_\odot}$ between two subpopulations with different $\chi_{\rm eff}$ distributions (in the primary-mass function), which is almost identical to $m_{\rm d}=46.1^{+5.6}_{-5.1}M_\odot$ found in the spin magnitude analysis by Wang et al. \cite{2022ApJ...941L..39W} with GWTC-3 data. 
Similar results are also presented in some other analysis based on GWTC-4.0 \cite{2025arXiv250915646B,2025arXiv250909876G,2025arXiv250909123A}.
While these new constraints are broadly consistent with earlier results mentioned above (and the cutoff at $\sim 45M_\odot$ has been attributed to the PISN explosions), as we will show in the following context, our analyses on the current/extended data with more flexible models and based on the hierarchical Bayesian inference reveals significant differences on the constraints for this critical mass (see also \cite{2025arXiv250923897L} for an earlier similar result, though 
it concentrates on the massive high-spin population). 

\begin{figure*}
	\centering  
\includegraphics[width=0.9\linewidth]{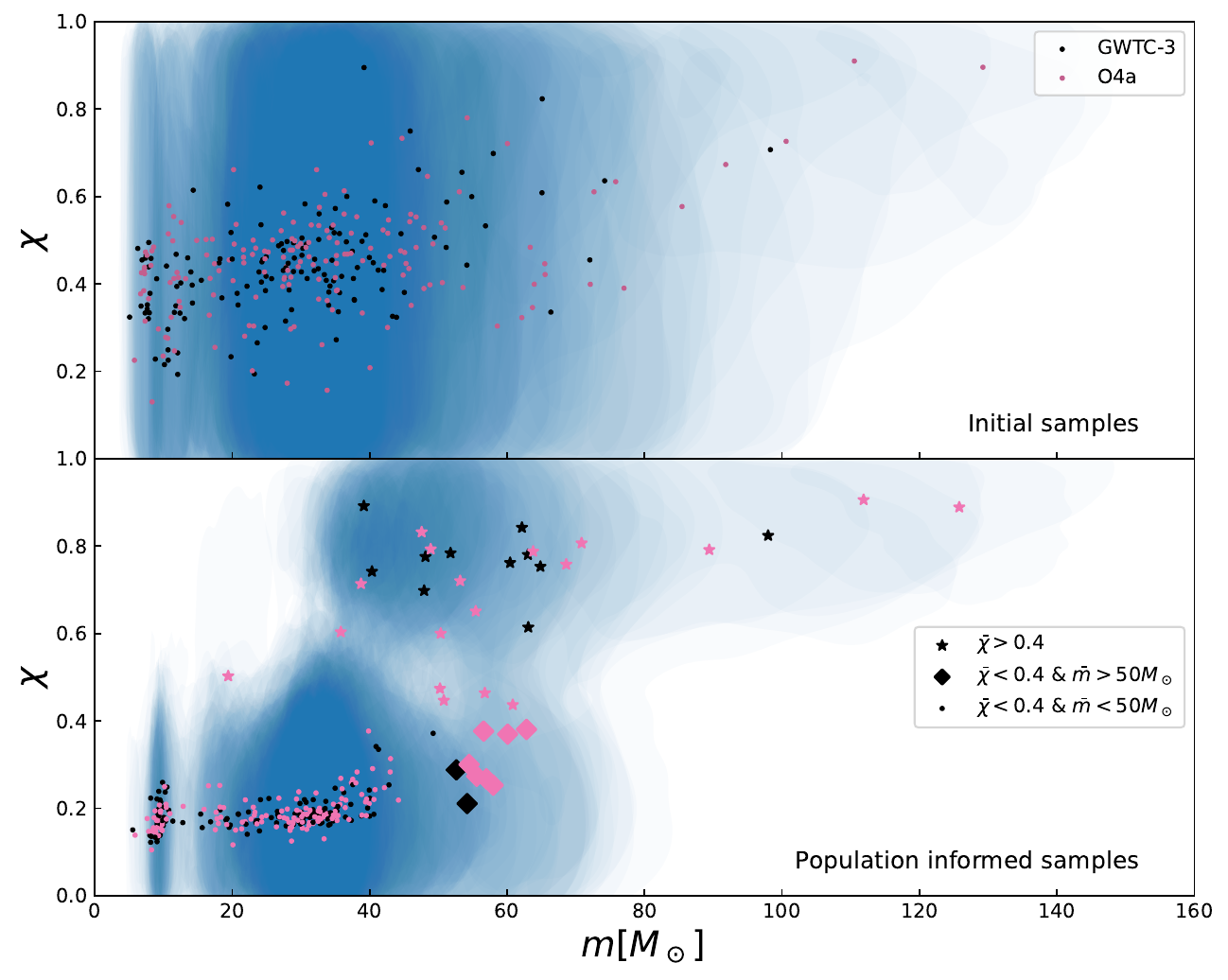}
\caption{The mass-spin distribution of the black holes in GWTC-4.0.
Top panel presents the original data,where the points (black for GWTC-3, while pink for O4a) are for the median values ($\bar{m}$,$\bar{\chi}$), and the shaded regions are for the 90\% credible intervals. 
An attractive feature we find is the emergence of a new group of $\gtrsim 50M_\odot$ black holes likely with low spins in the O4a data.
Bottom panel is for the {\it component-mass and spin-magnitude} (including both primary and secondary BHs) distributions of events in GWTC-4, reweighed by a population-informed prior inferred in this work. The points, diamonds, and stars are for the low-spin low-mass ($\bar{\chi}<0.4$, $\bar{m}<50M_\odot$), {low-spin high-mass} ($\bar{\chi}<0.4$, $\bar{m}>50M_\odot$), and high-spin ($\bar{\chi}>0.4$) BHs. } 
\label{fig:m1m2_dist}
\end{figure*}

As a start, to highlight the new information carried by the O4 data, we show the released posterior distributions of component-mass versus spin-magnitude of each event in GWTC-4.0 in the top panel of Figure \ref{fig:m1m2_dist}. The patterns of inferred medians for GWTC-3 events (black dots) and O4a events (pink dots) exhibit a discernible difference: the fraction of events that are more likely to have low component spins ($\chi < 0.5$) around $m \gtrsim 60M_\odot$ are larger. However, we caution that this is just a demonstration of the original data, and they are affected by the astrophysical distributions, the detector noise and the selection effects. If there is a new group of low-spin yet massive ($\gtrsim 60M_\odot$) BHs, they are difficult to produce via hierarchical mergers, and may instead point to either a low $^{12}{\rm C}(\alpha,\gamma)^{16}{\rm O}$ reaction rate \cite{2020ApJ...902L..36F,2021ApJ...912L..31W} or formation from massive stars whose envelopes were not stripped \cite{2024MNRAS.529.2980W}. To study how the new patterns in the data will affect the inferred population of black holes and their astrophysical implications, we carried out this work. In Sec.\ref{sec:Models} and Sec.\ref{sec:methods} we describe our population models and the framework of hierarchical Bayesian inference; in Sec.\ref{sec:Results} we analyse the inferred results and study the new constraints on the $^{12}{\rm C}(\alpha,\gamma)^{16}{\rm O}$ reaction rate; finally, in Sec.\ref{sec:Discussion} we summarise our findings and make some further discussions.

\section{Population Models}\label{sec:Models}
Our fiducial model describes the distribution of component masses, spin magnitudes, and spin orientations, which is developed from the mixture formula in Li et al. \cite{2024PhRvL.133e1401L}, see Section~\ref{main_model}.
For a cross check, we also carry out analysis with population model for distribution of the effective spins ($\chi_{\rm eff}\equiv \frac{m_1\chi_1\cos\theta_1+m_2\chi_2\cos\theta_2}{m_1+m_2}$), effective precession ($\chi_{\rm p}\equiv {\rm Max}\{\chi_1\sin\theta_1, \frac{(3m_1+4m_2)m_2}{(4m_1+3m_2)m_1}\chi_2\sin\theta_2\}$) and primary masses, see Section~\ref{alter_model}.

\subsection{The spin magnitude/orientation v.s. component mass model}\label{main_model}
In our fiducial model, the distribution in component mass and spin parameter space is taken as

\begin{equation}\label{twopop}
\begin{aligned}
&\pi(m,\chi,\cos\theta\mid\Lambda)= \\
&\quad \sum_{i=1,2} r_i P_{m,i}(m\mid\Lambda_i) P_{\chi,i}(\chi\mid\Lambda_i) P_{\cos\theta,i}(\cos\theta\mid\Lambda_i),
\end{aligned}
\end{equation}

where $i=1,2$ denotes there could be two different subpopulations of component BHs, and $r_i$ is the branch ratio of each subpopulation in the underlying distribution. Thanks to the significantly enriched data of GWTC-4, we adopt a comprehensive data-driven model, where all the component-mass, spin-magnitude and cosine-tilt-angle distributions ($P_{m,i}(m|\Lambda_i)$, $P_{\chi,i}(\chi|\Lambda_i)$, $P_{\cos\theta,i}(\chi|\Lambda_i)$) are described with cubic spline functions \cite{2023ApJ...946...16E}. The component-mass functions are expressed as  \cite{2022ApJ...924..101E,2023ApJ...946...16E}
\begin{equation}\label{eq:MF}
P_{m,i}(m|\Lambda)\propto m^{-\alpha_i} e^{f(m | \{f_j^i\}_{j=1}^{n})} {B}(m),
\end{equation}
where ${B}(m)$ is the function characterizing the boundary of the mass distribution defined as 
\begin{equation}
{B}(m)=
\begin{cases}
  0 & \text{if } m > m_{{\rm max},i} \\
  \mathcal{S} (m|m_{{\rm min},i}, \delta_{{\rm m},i})   &  \text{if } m_{{\rm min},i} < m < m_{{\rm max},i} \\
  0  & \text{if } m < m_{{\rm min},i}
\end{cases}
\end{equation}
$\mathcal{S} (m|m_{{\rm min},i}, \delta_{{\rm m},i})$ is the smooth function defined in \cite{2018ApJ...856..173T,2021ApJ...913L...7A}. For the first / second subpopulation, we set 15 / 10 knots ($\{f_j^1\}_{j=1}^{15}$) / ($\{f_j^2\}_{j=1}^{10})$) linearly in the log space between $[6-100]M_\odot$ / $[10-150]M_\odot$, considering the first (second) subpopulation dominates the lower (higher) mass range. Such a configuration is more flexible than that of previous work \cite{2024PhRvL.133e1401L,2025arXiv250923897L}, where 12 knots are use between $[6,80]M_\odot$. This enables us to explore the mass distributions of the two subpopulations in more details, especially for determining the locations of edges.
 
The spin magnitudes and cosine-tilt-angles are also described with flexible models \cite{2023ApJ...946...16E,2023PhRvD.108j3009G}
\begin{equation}
P_{\chi,i}(\chi|\Lambda)\propto e^{x(\chi | \{x_j^i\}_{j=1}^{n})}[\chi_{{\rm min},i},\chi_{{\rm max},i}],
\end{equation}
\begin{equation}
P_{\cos\theta,i}(\cos\theta|\Lambda)\propto e^{t(\cos\theta | \{t_j^i\}_{j=1}^{n})}[-1,1],
\end{equation}
we set 5 and 4 knots ($\{x_j^i\}_{j=1}^{5}$, $\{t_j^i\}_{j=1}^{4}$) to interpolate the perturbation function of spin magnitude and cosine-tilt-angle distributions in each subpopulation.

With the population model of component BHs defined above, then the overall population model for BBHs is expressed as 
\begin{equation}
\begin{aligned}
\pi(\lambda|\Lambda) \propto \pi&(m_1,\chi_1,\cos\theta_1|\Lambda)\pi(m_2,\chi_2,\cos\theta_2|\Lambda)
\\&
\times \mathcal{F}_{\rm pair}(m_1,m_2|\beta)P_z(z|\gamma),
\end{aligned}
\end{equation} 
where $\mathcal{F}_{\rm pair}(m_1,m_2|\beta)=(m_2/m_1)^\beta$ is the pairing function and $P_z(z|\gamma)$ is the redshift distribution, assuming a merger rate evolution $R(z)=R_0(1+z)^\gamma$. 
The parameters and priors for this fiducial population model are described in Table~\ref{tab:prior}.

\begin{figure*}
	\centering  
\includegraphics[width=0.9\linewidth]{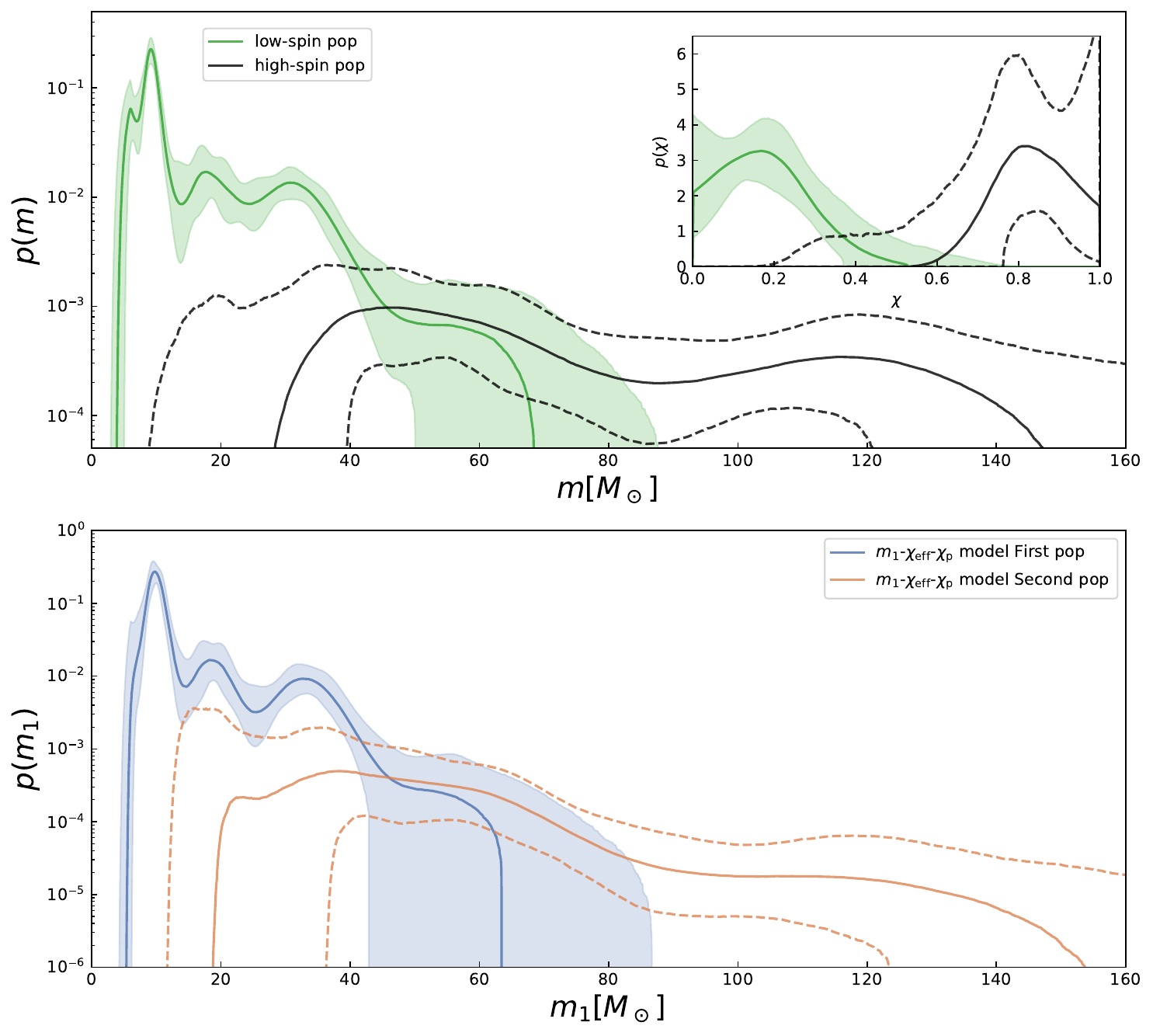}
\caption{The mass distributions of subpopulations. The solid lines (dashed lines / shaded regions) indicate the median values and $90\%$ credible regions for different subpopulations. Top panel presents the component-mass distribution inferred with the spin magnitude/orientation-mass model (fiducial model). 
Bottom panel is for the $m_1-\chi_{\rm eff}-\chi_{\rm p}$ model. In both cases, a cutoff at $\sim 60-70M_\odot$ is evident for the low-spin / first population.} 
\label{fig:mass_dist}
\end{figure*}

\begin{figure*}
	\centering  
\includegraphics[width=0.9\linewidth]{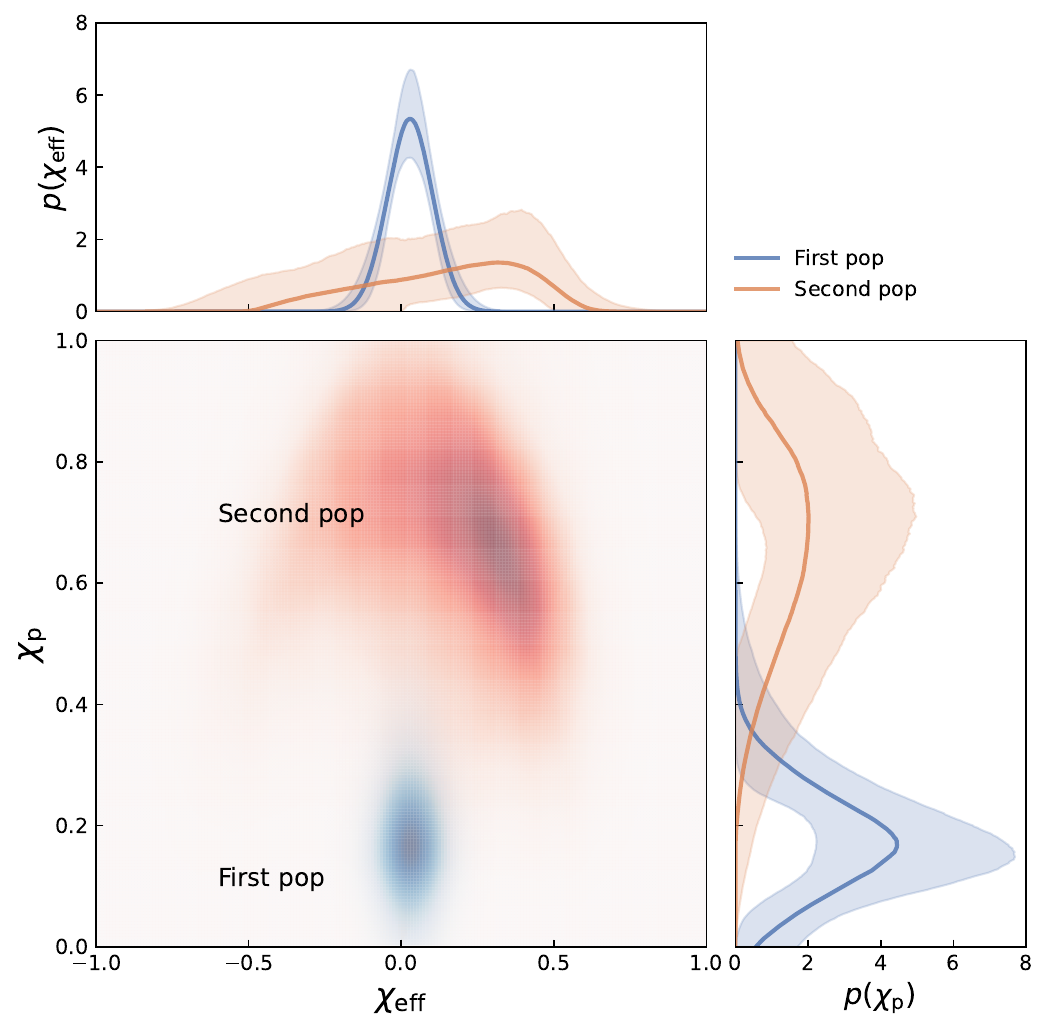}
\caption{The $\chi_{\rm eff}$ and $\chi_{\rm p}$ distributions for the two subpopulations inferred with the $m_1-\chi_{\rm eff}-\chi_{\rm p}$ model. The solid lines (dashed lines / shaded regions) indicate the median values and $90\%$ credible regions for different subpopulations.} 
\label{fig:X_dist}
\end{figure*}

\begin{figure*}
	\centering
\includegraphics[width=1.0\linewidth]{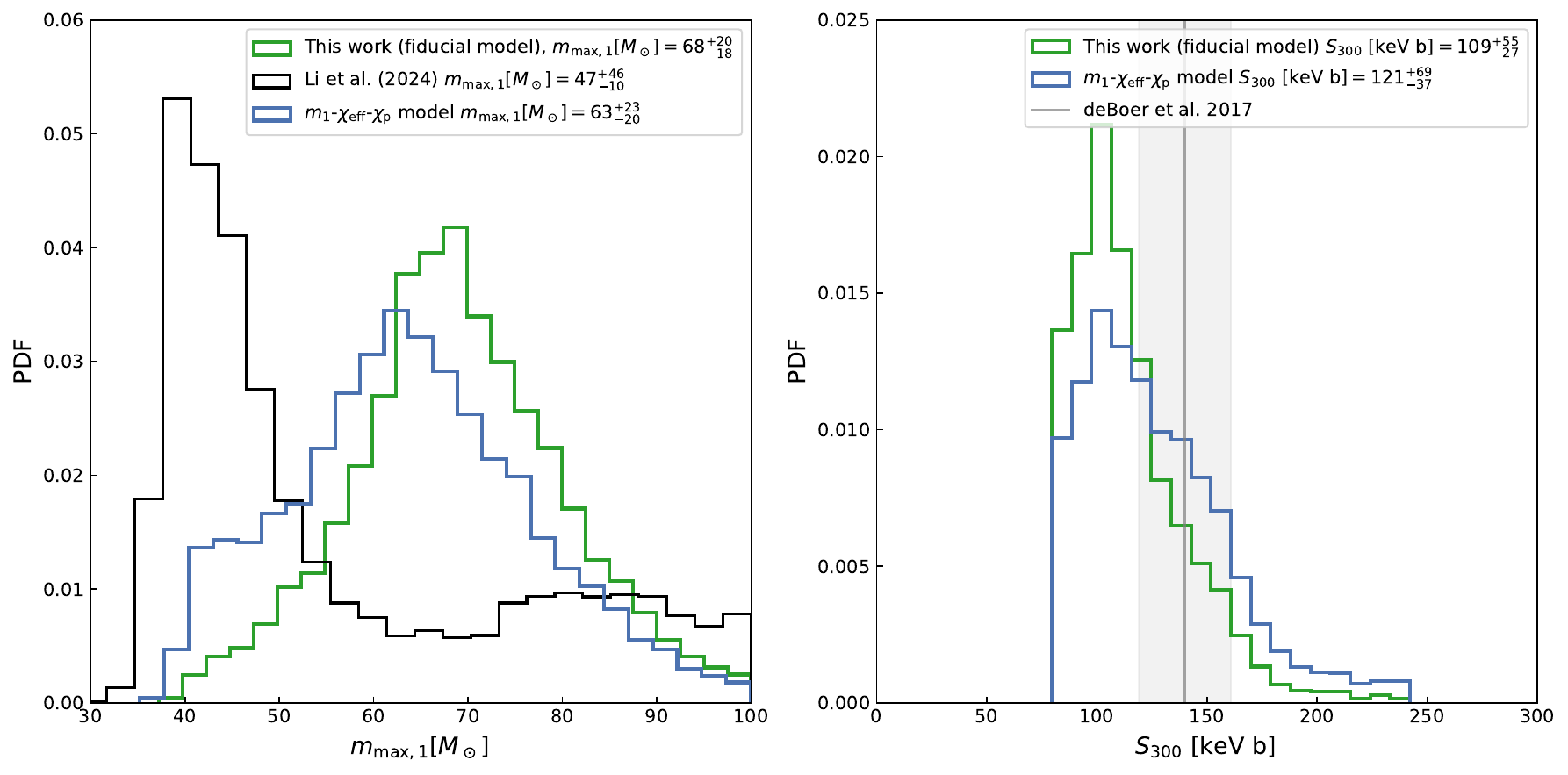}
\caption{The inferred $m_{\rm max,1}$ and the corresponding $S_{\rm 300}$. The left panel is the probability distribution of the maximum mass ($m_{\rm max,1}$) of the low-spin group, and we have $m_{\rm max,1}=68.5^{+19.8}_{-18.5}M_\odot$. In comparison to  \cite{2024PhRvL.133e1401L}, the peak of the current probability distribution of the maximum mass is significantly higher. 
The right panel is the evaluated $S$ factor of $^{12}{\rm C}(\alpha,\gamma)^{16}{\rm O}$ reaction at 300 keV ($S_{300}$) assuming that $m_{\rm max,1}$ represents $M_{\rm low}$. The value of $S_{300}$ recommended by \cite{2017RvMP...89c5007D} (1$\sigma$ region including the model uncertainty) is also plotted for comparison.}
\label{fig:S300}
\end{figure*}

\subsection{The $m_1-\chi_{\rm eff}-\chi_{\rm p}$ model}\label{alter_model}
{To cross check the inferred $m_{\rm max,1}$ (i.e., the upper mass cutoff of the low-spin / first subpopulation), we also develop a $m_1-\chi_{\rm eff}-\chi_{\rm p}$ model, which reads}
\begin{equation}
\begin{split}
\pi(m_1,\chi_{\rm eff},\chi_{\rm p},m_2,z|\Lambda) = & P_z(z|\gamma) P(m_2|m_1;m_{\rm min,1},\beta) 
\\ & \times 
\prod_{i=1,2}{P_i(m_1)P_i(\chi_{\rm eff},\chi_{\rm p})}r_i,
\end{split}
\end{equation}

where $r_i$ is the fraction of $i$-th subpopulation, $P_i(m_1)$ is the same as Eq.~(\ref{eq:MF}) defined in the fiducial model, and 
\begin{equation}
P(m_2|m_1;m_{\rm min,1},\beta) \propto {m_2^\beta}_{[ m_{\rm min,1}, m_1 ]}.
\end{equation}
The ($\chi_{\rm eff}$, $\chi_{\rm p}$) distribution for the first subpopulation (low-spin) can be simply expressed with Gaussian distributions, which reads
\begin{equation}
\begin{aligned}
P_1(\chi_{\rm eff},\chi_{\rm p}) \propto &\; \mathcal{N}(\chi_{\rm eff}|\mu_{\rm eff,1},\sigma_{\rm eff,1})_{[-1,1]} \\
& \times \mathcal{N}(\chi_{\rm p}|\mu_{\rm p,1},\sigma_{\rm p,1})_{[0,1]}.
\end{aligned}
\end{equation}
For the effective-spin and/or effective-precession distribution of the hierarchical mergers is predicted in many work \cite{2021PhRvD.104h4002B,2024A&A...685A..51V,2024ApJ...966L..16P,2025ApJ...990..217M,2025ApJ...993..163C}.  
We introduce a semi-parametric model for spin parameters, 
\begin{equation}\label{eq:chi}
\begin{aligned}
P_2(\chi_{\rm eff},\chi_{\rm p}) \propto &\; h(\phi(\chi_{\rm eff},\chi_{\rm p})) \\
& \times g(\rho(\chi_{\rm eff},\chi_{\rm p}))_{\chi_{\rm eff}\in[-1,1],\chi_{\rm p}\in[0,1]},
\end{aligned}
\end{equation}
where
$h(\phi) \propto e^{f(\phi)} \sin{\phi}_{[\phi_{\rm min},\phi_{\rm max}]}$ 
with 
$\phi(\chi_{\rm eff},\chi_{\rm p}) = \arctan{\frac{\chi_{\rm p}/a}{\chi_{\rm eff}/b}}$,
and $g(\rho)=\mathcal{N}(\rho | 1, \sigma_\rho)$ with 
$\rho(\chi_{\rm eff},\chi_{\rm p}) = \sqrt{({\chi_{\rm p}/a})^2+({\chi_{\rm eff}/b})^2}$. The $e^{f(\phi)}$ is represented with cubic spline function, interpolated with 6 knots linearly in $[0 ,\pi]$.
Note that, if we fix $\phi_{\rm min}=-\pi/2,\phi_{\rm max}=\pi/2$, and fix $f(\phi)$ as a constant, then the $\chi_{\rm eff}-\chi_{\rm p}$ distribution is a semi-annulus of elliptical shape with semi-axes $(a, b)$ and a width of $\sigma_\rho$, similar to the prediction of hierarchical mergers, see e.g. \cite{2024ApJ...966L..16P,2025PhRvL.134a1401A}. 
However, variable $\phi_{\rm min}$, $\phi_{\rm max}$, and $f(\phi)$ can generate an asymmetric $\chi_{\rm eff}$ distribution, which may be associated with the AGN disk channels \cite{2024A&A...685A..51V,2025ApJ...990..217M,2025ApJ...993..163C}. The parameters and priors specific to this population are also summarized in Table~\ref{tab:prior}.

\section{Hierarchical Bayesian Inference}\label{sec:methods}
We perform hierarchical Bayesian inference to constrain our model parameters. The likelihood is constructed assuming an inhomogeneous Poisson process. For a series of measurements of $N_{\rm obs}$ events $\vec{d}$, the likelihood for the hyper-parameters $\mathbf{\Lambda}$ can be inferred via \cite{2019MNRAS.486.1086M,2021ApJ...913L...7A,2023PhRvX..13a1048A}
\begin{equation}\label{eq:llh}
\begin{aligned}
\mathcal{L}(\vec{d}\mid \mathbf{\Lambda}) \propto N^{N_{\rm obs}}\exp(-N \eta(\mathbf{\Lambda}))\prod_{i}^{N_{\rm obs}}\frac{1}{n_i}\sum_{k}^{n_i}\frac{\pi(\lambda_{i}^k\mid \mathbf{\Lambda})}{\pi(\lambda_{i}^k\mid \varnothing)},
\end{aligned}
\end{equation}
where $N$ is the expected number of mergers during the observation period and $\eta(\mathbf{\Lambda})$ is the detection efficiency, $\pi(\lambda \mid \mathbf{\Lambda})$ is the population model (hyper-prior) defined in Section~\ref{sec:Models}. Following the procedures described in \cite{2025arXiv250818083T}, we use the injection campaign from \cite{2025arXiv250810638E} (https://zenodo.org/records/16740128) to estimate $\eta(\mathbf{\Lambda})$. The $n_i$ posterior samples for the $i$th event and the default prior $\pi(\lambda^k \mid \varnothing)$ are taken from the data releases of \cite{2024PhRvD.109b2001A, 2023PhRvX..13d1039A, 2025arXiv250818082T} (https://gwosc.org/eventapi/html/GWTC/). We constrain the total Monte Carlo integration uncertainty as $\ln\delta_{\rm tot}<1$ \cite{2025arXiv250818083T}, ensuring an accurate likelihood evaluation. We adopt the same detection threshold as \cite{2025arXiv250818083T} (false alarm rate ${\rm FAR}<1/{\rm yr}$), yielding 153 BBH events. Note that GW190814 is excluded from our analysis due to its unclear nature \cite{2020ApJ...896L..44A}. We use the {\sc Bilby} package \cite{2019ApJS..241...27A} with {\sc PyMultinest} sampler \cite{2016ascl.soft06005B} to obtain the Bayesian evidences and posteriors of the hyper-parameters for each model.

\begin{table*}[htpb]
\centering
\caption{Summary of model parameters.}\label{tab:prior}
\begin{tabular}{lcccc}
\hline
\hline
Parameter     &  Description & Prior \\
\hline
\hline
\multicolumn{3}{c}{\bf mass function}\\
$m_{{\rm min},i}[M_{\odot}]$   & The minimum mass & $U(2,50)$  \\
$m_{\rm max,2}[M_{\odot}]$ / $m_{\rm max,1}[M_{\odot}]$   & The maximum mass & $U(20,100)$ / $U(20,200)$  \\
$\alpha_i$ & Slope index of the power-law mass function & $U(-8,8)$ \\
$\delta_{\rm m,1}[M_{\odot}]$ / $\delta_{\rm m,2}[M_{\odot}]$ & Smooth scale of the mass lower edge & $U(0,10)$ /$U(0,20)$ \\
$\{f_j^1\}_{j=2}^{14}$ / $\{f_j^2\}_{j=2}^{9}$ & Interpolation values of perturbation function & $\mathcal{N}(0,1)$ \\
$r_2$ & mixture fraction for the second subpopulation & $U(0,1)$ \\
constraints & & $m_{{\rm min},i}<m_{{\rm max},i}$ \\
$\beta_{q}$ & Slope index of the mass-ratio distribution & $U(-8,8)$ \\
\hline
\multicolumn{3}{c}{\bf Spin distribution}\\
$\chi_{{\rm min},1}$ / $\chi_{{\rm min},2}$  & Lower edge for $\chi$ distribution  & 0 / $U(0,0.8)$  \\
$\chi_{{\rm max},1}$ / $\chi_{{\rm max},2}$  & Upper edge for $\chi$ distribution &  $U(0.2,1)$ / 1  \\
$\{x_j^i\}_{j=1}^{5}$ & Interpolation values for $\chi$ distribution & $\mathcal{N}(0,1)$ \\
$\{t_j^i\}_{j=1}^{4}$ & Interpolation values for $\cos\theta$ distribution  & $\mathcal{N}(0,1)$ \\
\hline
\multicolumn{3}{c}{\bf Rate evolution model} \\
$\lg (R_0[{\rm Gpc}^{-3}~{\rm yr}^{-1}])$ & Local merger rate  density & $ U(-3,3)$ \\
$\gamma$ & Slope of the power-law & $U(-8,8)$ \\
\hline
\multicolumn{3}{c}{\bf Special for $m_1-\chi_{\rm eff}-\chi_{\rm p}$ model}\\
$\phi_{\rm min}$  & Lower cut for $\phi$ defined in Eq.~(\ref{eq:chi})  &  $U(-\pi/2,\pi/2)$  \\
$\phi_{\rm max}$  & Upper cut for $\phi$ defined in Eq.~(\ref{eq:chi})  &  $U(-\pi/2,\pi/2)$  \\
constraints & & $\phi_{\rm min}<\phi_{\rm max}$ \\
$\{f_j\}_{j=1}^{6}$ & Interpolation values for $f(\phi)$ & $\mathcal{N}(0,1)$ \\
$a$ / $b$ & semi-axes for $\chi_{\rm p}$, $\chi_{\rm eff}$ & $U(0,1)$ / $U(0,1)$\\
$\mu_{\rm eff,1}$ / $\sigma_{\rm eff,1}$ & peak / width for $\chi_{\rm eff}$ distribution of first pop &  $U(0,1)$ / $LU(10^{-1},10^{0.5})$ \\
$\mu_{\rm p,1}$ / $\sigma_{\rm p,1}$ & peak / width for $\chi_{\rm p}$ distribution of first pop &  $U(0,1)$ / $LU(10^{-1},10^{0.5})$ \\
\hline
\hline
\end{tabular}
\\
\begin{tabular}{l}
Note: $U$, $LU$, and $\mathcal{N}$, represent Uniform, LogUniform, and Gaussian distribution respectively.
\end{tabular}
\end{table*}

\section{Results} \label{sec:Results}
Consistent with GWTC-3 studies \cite{2022ApJ...941L..39W,2024PhRvL.133e1401L,2025PhRvL.134a1401A}, we identify two subpopulations of BHs in GWTC-4.0, which are well separated in the component-mass versus spin-magnitude parameter space (see bottom panel of  Figure~\ref{fig:m1m2_dist}) and the $m_1-\chi_{\rm eff}-\chi_{\rm p}$ parameter space (see Figure~\ref{fig:mass_dist} and Figure~\ref{fig:X_dist}). 
The second subpopulation's spin magnitudes peak at $\sim 0.8$ and its mass distribution extends beyond $100 M_\odot$, and we refer the readers to  Li et al. \cite{2025arXiv250923897L} for the dedicated investigation on the high-spin subpopulation with a similar population model. These features are most naturally explained if these BHs are remnants of earlier mergers \cite{2021NatAs...5..749G}, especially for aligned BBH mergers in active galactic nucleus disks \cite{2024A&A...685A..51V,2025arXiv250923897L}. Additionally, some events may contain double high-spin BHs, as was also suggested in \cite{2020PhRvL.125j1102A,2025arXiv250717551L,2025arXiv250904706A,2025arXiv250908298L}.

In the left panel of Figure \ref{fig:S300}, we plot the posterior distribution of the mass cutoff for the low-spin BH population that can be naturally attributed to the stellar-collapse origin and find $m_{\rm max,1}=68.5^{+19.9}_{-18.3}M_\odot$ ($90\%$ credibility; see also  \cite{2025arXiv250923897L} for a similar result). The metallicity-dependent stellar winds, convective overshooting, internal mixing and stellar rotation can shift the PISN boundary, but the $^{12}{\rm C}(\alpha,\gamma)^{16}{\rm O}$ reaction rate is a key uncertainty in setting $M_{\rm low}$ \cite{2019ApJ...887...53F}. 
Our inferred mass cutoff, although higher than the $M_{\rm low}$ predicted in many works \cite{2016A&A...594A..97B, 2017MNRAS.470.4739S, 2017ApJ...836..244W, 2023ApJ...945...41S}, is possible for a relatively low $^{12}{\rm C}(\alpha,\gamma)^{16}{\rm O}$ reaction rate \cite{2021ApJ...912L..31W,2022ApJ...924...39M} or even a normal one \cite{2026RAA....26g5011X}. 
{Combining the $M_{\rm BH}$-$\sigma$ relation from Figure~10 (where $\sigma$ is the uncertainty in the $^{12}{\rm C}(\alpha,\gamma)^{16}{\rm O}$ reaction rate) of Mehta et al. \cite{2022ApJ...924...39M} with the reaction rate from deBoer et al. \cite{2017RvMP...89c5007D}},
we infer $S_{\rm 300}=108.6^{+54.9}_{-26.5}~{\rm keV~b}$ ($90\%$ confidence interval; see Figure \ref{fig:S300}), potentially lower than the commonly adopted $S_{\rm 300}=140\pm 21^{+18}_{-11}~{\rm keV~b}$ recommended by deBoer et al. \cite{2017RvMP...89c5007D}. Note that with new measurements of the ground state asymptotic normalization coefficient, Shen et al. \cite{2023ApJ...945...41S} obtained a higher value of $S_{\rm 300}\approx 170~{\rm keV~b}$ \footnote{While this finding has been widely cited in the literature to support a scenario with a high 
$S_{300}$ value, a critical caveat must be noted: the authors arrived at this elevated 
$S_{300}$ by employing an ``enhanced" 
$S(E2)$ component while fixing the 
$S(E1)$ term in their calculations. However, their more recent work \cite{2024PhRvC.109d5808N} reports a revised 
$S(E1)$ value of 
$55~{\rm keV~b}$, a significant reduction from the previously fixed parameter. When this updated 
$S(E1)$ is incorporated into the analysis, the resulting 
$S_{300}$ aligns closely with the benchmark value recommended by deBoer et al. \cite{2017RvMP...89c5007D}.}, which suggests a $M_{\rm low}\approx 52M_\odot$, plausibly in tension with our $m_{\rm max,1}$. However, a recent Bayesian analysis favors a value of $S_{\rm 300}\approx 130~{\rm keV~b}$ \cite{2025arXiv250917102M}.
The BHs near this cutoff have spin magnitudes of $\chi\lesssim0.4$, broadly continuous with the lower-mass stellar-origin population. We therefore interpret them as the high-mass extension of stellar-origin BHs made possible by an upward-shifted $M_{\rm low}$. It is thus reasonable to associate $m_{\rm max,1}$ with the upper end of the stellar-origin BH birth-mass spectrum, i.e. $M_{\rm low}$ in the PISN scenario.

Identifying a population of low-spin, high-mass BHs initially formed in massive single-star evolution would also establish the dynamical capture channel for BBH formation. With a larger sample anticipated in the upcoming O5 run, subtracting this component could yield a more precise estimate of $M_{\rm low}$ (which is $\approx m_{\rm max,1}$) and thus $S_{\rm 300}$. 
Hendriks et al. \cite{2023MNRAS.526.4130H} showed that shifting the expected carbon-oxygen core mass range upward by $\sim 10M_\odot$ brings the predicted rate of hydrogen-poor superluminous supernovae into better agreement with observations.
In such a scenario, they predicted no cutoff in the BH mass function up to $\sim 64M_\odot$, which may have been confirmed by this work.

With the enlarged GWTC-4.0 sample, we recover a 
low-spin yet high-mass BH group extending to $\gtrsim 60-70\,M_\odot$, 
whose properties are difficult to attribute to hierarchical growth \cite{2017PhRvD..95l4046G,2017ApJ...840L..24F,2021NatAs...5..749G}.
One potential interpretation for this low-spin, high-mass black hole group is the collapse of massive single stars \cite{2021ApJ...912L..31W}. 
It was realized by Winch et al. \cite{2024MNRAS.529.2980W} that blue supergiant progenitors with small cores but large hydrogen envelopes at low metallicity could directly collapse to black holes as massive as $\sim 93M_\odot$. 
Such massive first-generation BHs (with low spins) can be assembled via dynamical capture in dense environments \cite{2021NatAs...5..749G}. In this scenario, the spins of these BHs would be isotropic, unless being oriented in gas-rich environments \cite{2024MNRAS.531.3479M}.
Unfortunately, the current sample of low-spin, high-mass BH events is too small to test this expectation, 
and analyzing the spin orientation distribution for low-spin events is inherently challenging \cite{2025PhRvD.112h3015V}. If the mass of these objects were instead inherited from the CO core of massive stars, they raise the lower edge of the PIMG to $M_{\rm low}=68.5^{+19.8}_{-18.3}M_\odot$, and implies $S_{\rm 300}=108.6^{+54.9}_{-26.5}~{\rm keV~b}$. 
Note that if we instead adopt the theoretical 
$M_{\rm low}-S_{\rm 300}$
  curve presented by Xin et al. \cite{2026RAA....26g5011X}, the derived 
$S_{\rm 300}$ value will be in excellent agreement with the benchmark value recommended in \cite{2017RvMP...89c5007D}.

The above results are obtained in our fiducial spin magnitude/orientation-mass distribution model. For the $m_1-\chi_{\rm eff}-\chi_{\rm p}$ distribution model, the resulting $m_{\rm max,1}$ and $S_{300}$ are rather similar (see Figure \ref{fig:S300}), demonstrating the robustness of our conclusions. Moreover,
Figure~\ref{fig:X_dist} shows that the $\chi_{\rm eff}$ distribution is asymmetric, supporting the conclusion made with GWTC-3 \cite{2025ApJ...987...65L} and the results inferred with component spins in GWTC-4 \cite{2025arXiv250923897L}.
It is notable that the minimum mass of the second subpopulation can be lower than $20M_\odot$, which is also possible in the fiducial model (see Figure \ref{fig:mass_dist} and the original discussion in \cite{2025arXiv250923897L}).  There is likely a peak at $\sim18M_\odot$, consistent with the masses of remnants from BBHs at $\sim10M_{\odot}$. This indicates that the BBHs at $\sim10M_{\odot}$ may also contribute to the dynamical formation channels.

\begin{figure*}
\centering  
\includegraphics[width=0.9\linewidth]{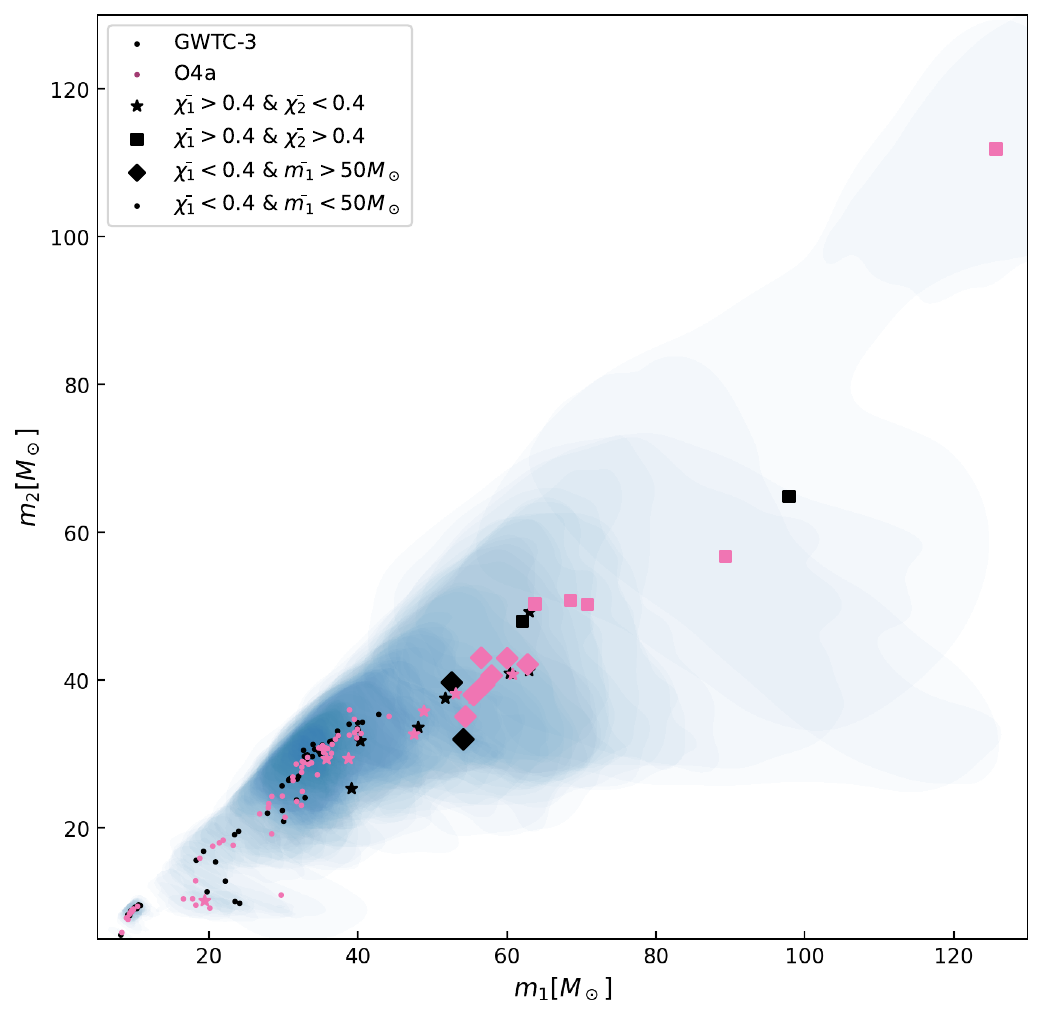}
\caption{Primary-mass versus secondary-mass distributions of events in GWTC-3 and O4a, reweighed by a population-informed prior inferred in this work. The points, diamonds, and stars / squares are for the low-spin low-mass ($\bar{\chi_1}<0.4$, $\bar{m_1}<50M_\odot$), {low-spin high-mass} ($\bar{\chi_1}<0.4$, $\bar{m_1}>50M_\odot$), and high-spin ($\bar{\chi_1}>0.4$, $\bar{\chi_2}<0.4$) / ($\bar{\chi_1}>0.4$, $\bar{\chi_2}>0.4$) BBHs. } 
\label{fig:ma_dist}
\end{figure*}

\section{Discussions.}\label{sec:Discussion}
It is the time to understand the difference of the cutoff masses reported for the first-generation population in some recent literature. Tong et al. \cite{2025arXiv250904151T} focused on the mass function of the secondary BHs (assumed to trace stellar-origin black holes) and, within that framework, they reported a cutoff at $\approx 45M_\odot$. However, their assumption may break down if the secondary component is a mixture that includes higher-generation secondaries; for this possibility, see the square markers in Fig.~\ref{fig:ma_dist} from our analysis. 
Very recently, with a flexible model, Ray \& Kalogera \cite{2025arXiv251018867R}
argued that the $m_2$ cutoff at $\sim 40-50M_\odot$ reported in \cite{2025arXiv250904151T} may be caused by the strong prior assumptions, and the intrinsic value should be higher. 
In our analysis, there is no good candidate for low-spin secondary BHs above $\sim 50M_\odot$ (note that some high-spin objects are heavier). But this is likely just a coincidence, since for a $m_{\rm max,1}\approx 68M_{\odot}$ and a typical $q\sim 0.7$ (see Figure~\ref{fig:ma_dist}),  the mass function of the secondary BHs will get suppressed at $\geq q m_{\rm max,1}\sim 45M_\odot$.  So, at least in our scenario, it is improper to interpret the lack of low-spin $m_{2}$ at $\geq 45M_{\odot}$ as the low edge of the PIMG.
Antonini et al. \cite{2025arXiv250904637A} reported a rapid effective-spin transition at $45.3^{+6.5}_{-4.8} {M_\odot}$ and interpreted it as $M_{\rm low}$,  in agreement with the spin-magnitude-based analysis result by
Wang et al. \cite{2022ApJ...941L..39W} in 2022 with the GWTC-3 data. We note that in Figure \ref{fig:mass_dist}, the declining mass function of the low-spin population intersects the rising mass function of the high-spin population at $\approx 45M_\odot$. This can explain why, in the analyses assuming a mass-dependent spin-magnitude or effective spin distribution to model a transition between the two subpopulations, the authors obtained a dividing mass of $\approx 45M_\odot$ \cite{2022ApJ...941L..39W, 2025arXiv250904637A}. 
However, this mass is not necessary to mark the onset of PIMG and a higher value is plausible. 
We have further illustrated that the difference between our results and those of Antonini et al. \cite{2025arXiv250904637A} is not caused by the different spin parameters we used, as  a similar $m_{\rm max,1}=63^{+23}_{-20}M_\odot$ is inferred for the first subpopulation with the $m_1-\chi_{\rm eff}-\chi_{\rm p}$ model (see the left panel of Figure \ref{fig:S300}). This indicates that the onset of PIMG may be higher than $60M_\odot$, although lower values $\lesssim 50M_\odot$ cannot be convincingly ruled out with current data.

The catalog of detected binary black hole (BBH) events is expanding. 
The BBH sample is expected to be doubled, when all events of the O4 run have been released. At that time the properties of the first-generation BHs as well as the stellar evolution theories will be better revealed/probed. The prospect is even more promising since LIGO/Virgo/KAGRA will be further upgraded.\\

\begin{acknowledgments}
This work is supported in part by the National Natural Science Foundation of China (No. 12233011, No. 12588101, No. 12503059, No. 12203101, No. 12303056),  the New Cornerstone Science Foundation through the XPLORER PRIZE, the General Fund (No. 2024M753495) of the China Postdoctoral Science Foundation, and the Priority Research Program of the Chinese Academy of Sciences (No. XDB0550400). This research has made use of data and software obtained from the Gravitational Wave Open Science Center (https://www.gw-openscience.org), a service of LIGO Laboratory, the LIGO Scientific Collaboration and the Virgo Collaboration. LIGO is funded by the U.S. National Science Foundation. Virgo is funded by the French Centre National de Recherche Scientifique (CNRS), the Italian Istituto Nazionale della Fisica Nucleare (INFN) and the Dutch Nikhef, with contributions by Polish and Hungarian institutes.
\end{acknowledgments}

\medskip

\bibliographystyle{apsrev4-1}
\bibliography{export-bibtex}

\end{document}